# Blockchain Enabled Enhanced IoT Ecosystem Security

Mahdi H. Miraz[1] and Maaruf Ali[2]

[1] The Chinese University of Hong Kong, Shatin, NT, Hong Kong
[2] International Association of Educators and Researchers (IAER), London, UK
`m.miraz@cuhk.edu.hk, maaruf@ieee.org`

**Abstract.** Blockchain (BC), the technology behind the Bitcoin crypto-currency system - is starting to be adopted for ensuring enhanced security and privacy in the Internet of Things (IoT) ecosystem. Fervent research is currently being focused in both academia and industry in this domain. Proof-of-Work (PoW), a cryptographic puzzle, plays a vital rôle in ensuring BC security by maintaining a digital ledger of transactions, which are considered to be incorruptible. Furthermore, BC uses a changeable Public Key (PK) to record the identity of users – thus providing an extra layer of privacy. Not only in crypto-currency has the successful adoption of the BC been implemented, but also in multifaceted non-monetary systems, such as in: distributed storage systems, proof-of-location and healthcare. Recent research articles and projects/applications were surveyed to assess the implementation of the BC for IoT Security and identify associated challenges and propose solutions for BC enabled enhanced security for the IoT ecosystem.

## 1 Introduction

The primary aim of this article is to investigate the research question, "To what extent can the Blockchain be used in enhancing the overall security of the Internet of Things (IoT) ecosystems?" and to draw appropriate conclusions. Considering the fact that the Blockchain is comparatively an avant-garde technology, this paper presents a representative sample of research conducted in the last ten years, commencing with the early work in this domain. Although, identifying how the Blockchain can further enhance the security paradigm of IoT is the main focus of the paper, to do so various other usages of the Blockchain and similar digital ledger technologies were explored along with their applications, impediments, privacy and security concerns.

Like many other domains of computing, security and privacy issues are the major concerns of the Internet of Things (IoT) eco-system. To fortify the backbone for im-proved security and privacy of IoT, the Blockchain is considered to be able to play a vital rôle. In fact, Blockchain research has become truly multifaceted as researchers from both industry as well as academia are applying the Blockchain in new dimensions on a regular basis. In the Proof-of-Work (PoW) concept, as shown in Fig. 1, an algorithm based on mainly solving a mathematical challenge, is the major method to assure the security aspects of the BC by recording and maintaining a complete digital ledger of all the completed transactions. These transactions are thus unalterable.

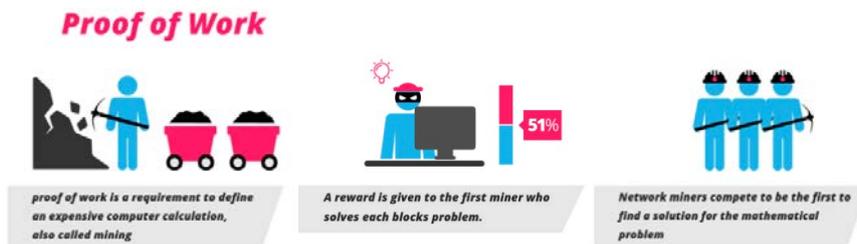

Fig. 1. The Proof of Work Concept [1].

A high-level system block diagram of how the BC technology works is shown in Fig. 2.

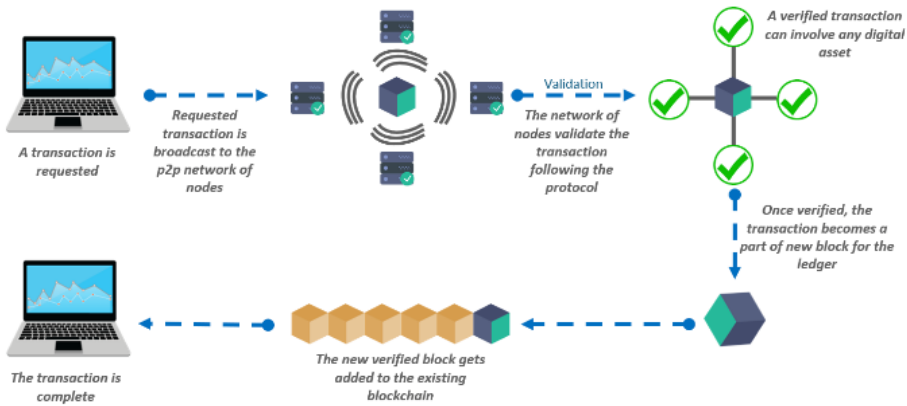

Fig. 2. A high level view of the Blockchain Technology [2].

In addition to this, the BC also takes advantage of the Public Key, as shown in Fig. 3, which is purposely made chaotic in nature for ensuring the highest level of security, in order to register the identity of the users. Thus, an extra layer of privacy is ensured automatically. As evident by many research and project reports, the adoption of the Blockchain technology has been found to be successful in many non-monetary domains such as in the supply chain, healthcare systems, online/electronic voting, proof of location, distributed cloud storage, even in human resource management and recruitment [3].

The authors of this paper not only surveyed research articles but also considered relevant projects/applications to ascertain the applicability of Blockchain technology for augmented IoT security and to distinguish the challenges associated with such application of the BC and thence to put forward probable solutions for BC enabled enhanced IoT security systems.

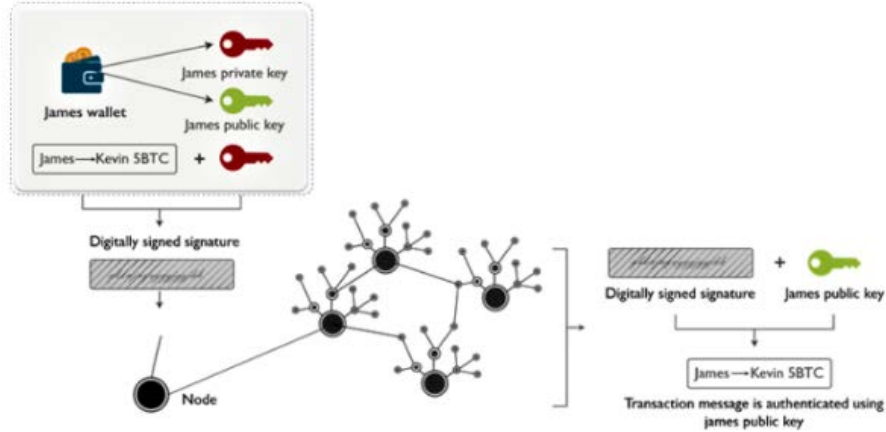

Fig. 3. The verification of signatures by the miners in the blockchain [2].

The knowledge domain of the research is in the realm of the Internet of Things (IoT), Internet of Everything (IoE), Wireless Sensor Network (WSN), digital ledger, specifically, in Blockchain and crypto-currency.

## 2    Blockchain Fundamentals

To understand how Blockchain can be applied for enhancing IoT security, it is very important to understand how these two technologies are put into function. In this section, the basic technological fundamentals of Blockchain have been briefly described while the next session introduces the Internet of Things (IoT) ecosystem.

A Blockchain mainly consists of two separate but interrelated integrant. These are as follows:
1. **Transaction:** in a digital ledger system such as the Blockchain, a transaction is basically the action triggered by the participant.
2. **Block:** A block, in a Blockchain system, is a collection or pool of data which records the transaction and other relevant details such as the correct sequence, timestamp of creation, *et cetera*.

Based on the scope of how a Blockchain is going to be used, it can be of two types: private or public. In a public Blockchain, usually all the users have both read and write permissions. One example of such a public Blockchain application is in recording the generation and financial flow of the Bitcoin cryptocurrency. However, there are also some public Blockchains where access is limited to either write or read rights, depending on the rôle of the user in the system. The aim of a private Blockchain, on the contrary, is to conceal the details of the users. To ensure that, access is limited to some trusted participants or members of a single organization. A Blockchain that is controlled by a consortium is known as a consortium blockchain. This is particularly pertinent amongst governmental institutions and allied sister concerns or their subsidies thereof.

The implementation of the Blockchain technology being public puts the BC well ahead of other technologies, especially in terms of security aspects. Since each participating nodes possesses its own copy of the complete blockchain i.e. whole blocks of updated records and transactions, the data thus remain unaltered. Any unauthorised or unexpected changes will thus be publicly verifiable. However, the data recorded in such publicly available blocks are hashed and encrypted (by the private key) to ensure security and anonymity. Because the private key is used to encrypt the data, it cannot be publicly interpreted, as shown in Fig. 3.

Although a centralised implementation of the Blockchain technology is possible, it is mostly decentralized in nature, which is considered to be another one of its major advantages. It is decentralized in the sense that:

- The data, comprising the transactions and associated blocks, are distributed among the participating nodes of the Blockchain network, rather than storing them in a single piece of node or storage device.
- The transactions are approved by a set of specific rules or algorithms, thus eliminating the influence of being biased by one single authority involving substantial trust in order to reach a consensus.
- The Blockchain systems only allows new verified blocks be appended to the old chain. As the previously added blocks are already public and distributed, they are openly verifiable and hence cannot be altered or revised. Thus, the overall security of a Blockchain ecosystem is another advantage over other technologies.

Once a transaction is triggered by a participant, it is not added straightaway to the chain of blocks i.e. the blockchain. In order for a newly initiated transaction to be appended with the existing chain, the transaction has to go through the validation and verification processes. The participating nodes of the Blockchain networks must apply a set of predefined rules or specific algorithms for this purpose. The set of rules or algorithms basically defines what is perceived as "valid" by the respective Blockchain system and may vary from one to another. Rather than adding one single transaction in a block, usually a number of such transactions are put together in order to construct a new block. This newly prepared block is then sent to all other participating nodes of the Blockchain network so that they can be appended to their copy of the existing chain of blocks. Each succeeding block of the chain comprises a hash, a unique digital fingerprint, of the preceding one.

The Blockchain not only verifies and validates all the newly triggered transactions but also maintains an irreversible lifelong record of them, while assuring that all the identification related information of the users or the participants are kept incognito. Thus all the personal information of the users is sequestered while substantiating all the transactions. This is achieved by reconciling mass collaboration by cumulating all the transactions in a computer code based digital ledger. Thus, in a Blockchain system, instead of trusting each other or an intermediary, the participants need to trust the decentralized network system itself. Thus the Blockchain itself has become the ideal "Trust Machine" [4,5] paradigm.

Although the Bitcoin cryptocurrency first used the Blockchain, it is considered to be just an exemplary use of the BC. Blockchain technology is a relatively novel technology in the domain of computing that is enabling illimitable applications, such as in and not just limited to: healthcare systems, human resource management, recruitment, storing and verifying legal documents including deeds and various certificates, IoT and the Cloud. In fact, Tapscott [6] has perfectly connoted Blockchain to be the "World Wide Ledger", facilitating many novel applications beyond just the simple verifying of transactions such as in: recording smart deeds, decentralized and/or autonomous organizations/government services *et cetera*. Fig. 4, shows the typical and diverse applications of the blockchain technology.

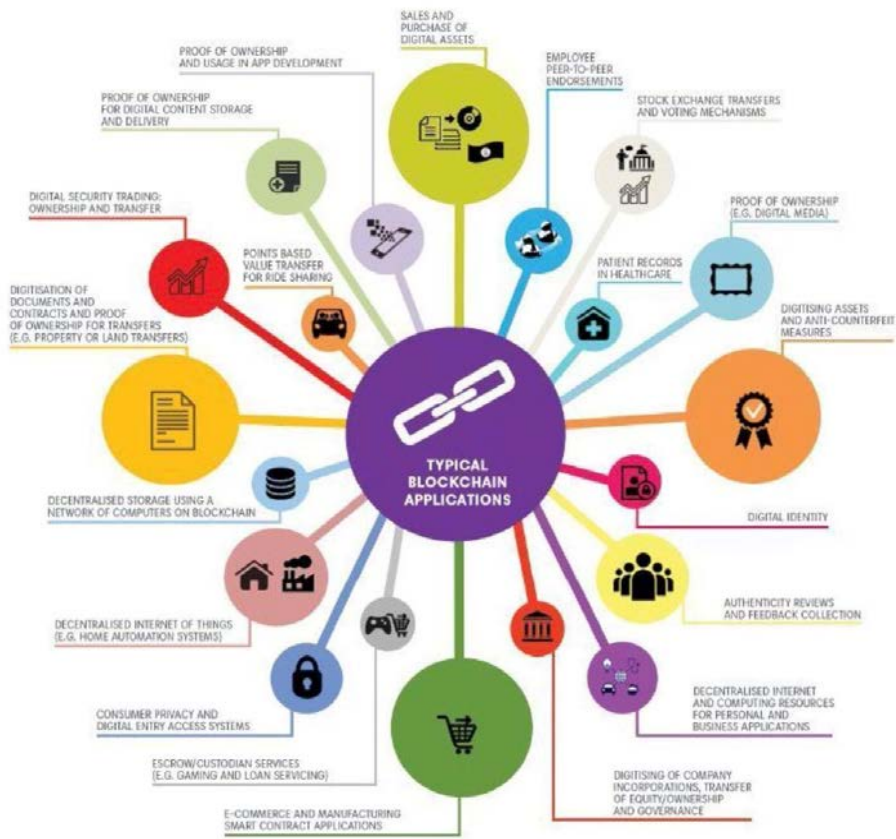

Fig. 4. Typical Application of the Blockchain Technology [7].

## 3      Internet of Things (IoT)

The term 'Internet of Objects' or 'Internet of Things' (more commonly referred to as 'IoT') - denotes the electronic or electrical devices of many different sizes and capabilities connected to the Internet. This connection is mainly by using wireless sensors, but excluding those primarily involved in communications with human beings, i.e. the traditional Internet. New IoT devices are being marketed on a regular basis and thus the scope of the connections is ever broadening beyond just basic machine-to-machine communication (M2M) [8].

There are many types of IoT devices employing a wide range of applications, protocols, and network domains [9]. The growing preponderance of IoT technology is enabled by the physical objects being connected to the Internet by various types of short-range wireless technologies such as sensor networks, RFID, ZigBee and through location-based technologies [10].

The emergence of IoT as a distinctive entity was reached (according to the Internet Business Solutions Group (IBSG)) when more inanimate objects were directly connected to the Internet bypassing human users [11]. This accelerating process has been gaining momentum ever since the rollout of CISCO's 'Planetary Skin', the Smart Grid and intelligent vehicles [11]. IoT is already on the verge of making the Internet truly pervasive, with devices already embedded into consumer white goods, including personal and intimate devices in our daily lives. IoT devices are only standardised in their use of the Internet networking protocols and not how they interface to the Internet or with each other. This immediate potential inhibiting factor needs to be addressed.

IoT may be deployed with added privacy, security and management features to link, for example, vehicle electronics, home environmental management systems, telephone networks and control of domestic utility services. The broadening scope of IoT and how it can link with heterogeneous networks is shown in Fig. 5 [11], below.

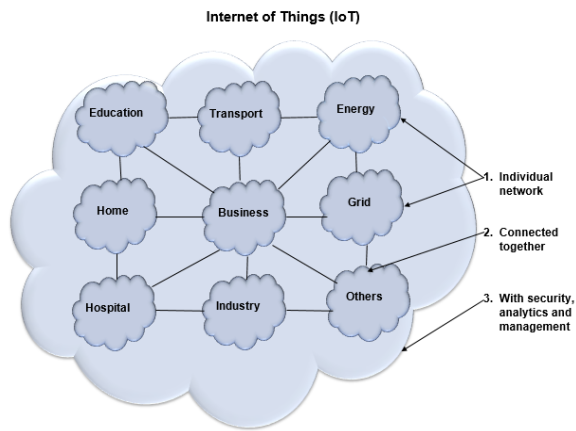

**Fig. 5.** IoT viewed as a Network of Networks [11].

A standard IoT ecosystem typically comprises of the following five components:
1. Sensors: sensors are mainly responsible to collect and transduce the required data;
2. Computing Node: such nodes containing the central processing unit (CPU), are required for processing the data and information received from a sensor;
3. Receiver: which is actually a transceiver, facilitates the collection of the message sent by the local and remote computing nodes or other associated devices;
4. Actuator: which could be electro-mechanical in nature, works on the basis of the decision taken by the Computing Node, processing the information received from the sensor and/or from the Internet, then triggering the associated device to perform a function; and
5. Device: to perform the desired task as and when triggered.

## 4    BC Enabled Enhanced IoT Security

In an IoT ecosystem [12,13], most of the communication is in the form of Machine-to-Machine (M2M) interactions, that is, without any human intervention whatsoever. Thus establishing trust among the participating machines is a big challenge that IoT technology still has not met extensively. However, Blockchain may act as a catalyst in this regard, by enabling enhanced scalability, security, reliability and privacy [4,5]. This can be achieved by deploying Blockchain technology to track billions of devices connected to the IoT ecosystems and then used to enable and/or coordinate transaction processing. In fact, a specific search engine exists, called "Shodan", that describes itself as "the world's first search engine for Internet-connected devices" [14]. The use of this search engine by anyone will also expose any insecure IoT devices and hence their need for rectification. Application of the Blockchain in any IoT ecosystem will further enhance the reliability by completely eliminating any Single Point of Failure (SPF). In Blockchain, data is encrypted using cryptographic algorithms as well as the hashing techniques. Thus, the application of Blockchain in an IoT ecosystem can offer better security services. However, to perform the hashing techniques and implement the cryptographic algorithms, the systems shall obviously demand more processing power, which IoT devices currently lack. Thus further research is required to overcome this present limitation, including extending longevity of the powering source.

Underwood [15] considers the application of Blockchain technology to completely overhaul the digital economy. Ensuring and maintaining trust is both the primary and initial concern of the application of the blockchain. BC can also be used to gather chronological and sequence information of transactions, as it may be seen as an enormous networked time-stamping system. For example, NASDAQ is using its 'Linq blockchain' to record its private securities transactions. Meanwhile the Depository Trust & Clearing Corporation (DTCC, USA) is working with Axoni in implementing financial settlement services such as post-trade matters and swaps. Regula-tors are also interested in BC's ability to offer secure, private, traceable real-time monitoring of transactions.

Securing operational technology is also of paramount importance. Thus the Blockchain can help to prevent tampering and spoofing of data by managing and securing industrial IoT and operational technology (OT) devices. So once a sensor, device or controller has been deployed and is working, it cannot be touched. Since any compromised devices will be recorded in the BC. Thus as Pindar propounds:

## 5    Concluding Discussions

To answer the research question "To what extent can the Blockchain be used in enhancing the overall security of the Internet of Things (IoT) ecosystems", this paper first introduced how these two emerging technologies works. The current security issues related to IoT systems were also discussed. The authors of the article then investigated how the application of the Blockchain can eliminate these security concerns inherent in the IoT ecosystem and improve its overall security.

Both the Blockchain and Internet of Things (IoT) are two relatively new but promising technologies being successfully used in multifaceted applications. The way the application of the Blockchain has widened beyond its initial use for Bitcoin generation and dealing has conclusively shown its relevance and versatility in general networked secure transactions. IoT also proved itself to be capable of doing far more things than being a simple wireless sensor network. In fact, Blockchain and its variants combined offers many security aspects such as enhanced privacy, stronger security, full traceability, inherent detailed data provenance and accurate time-stamping which other technologies still could not offer as standalone features. Thus BC has seen its adoption beyond its initial application areas and is now used to secure any type of transactions, whether: human-to-human, machine-to-machine or human-to-machine communications. The adoption of Blockchain appears to be secure, especially allied with the world emergence of the Internet-of-Things (IoT). Its decentralized application across the already established global Internet is also very appealing, in terms of ensuring network redundancy, data redundancy through distribution and hence survivability.